\documentclass[11pt]{article}
\usepackage{url,epsfig}
\usepackage[colorlinks]{hyperref}
\usepackage{fancyhdr,psfig,lastpage}
\usepackage{amsmath}
\usepackage{latexsym}
\usepackage[sort&compress,numbers]{natbib}
\begin{document}

\vspace*{0.8cm}
\begin{center}
  {\Large \bf Self-Organization, Active Brownian Dynamics, \\[2mm] 
and Biological Applications}\\[10mm]
 
{\large Werner Ebeling$^{1}$, Frank Schweitzer$^{1,2}$}\\[3mm]

\begin{quote}
\begin{itemize}
\item[$^{1}$] \emph{Institute of Physics, Humboldt University, 
    Invalidenstra{\ss}e 110, 10115 Berlin, Germany}
\item[$^{2}$] \emph{Fraunhofer Institute for Autonomous Intelligent
    Systems, Schloss Birlinghoven, 53754 Sankt Augustin, Germany,  \texttt{schweitzer@ais.fhg.de} }
\end{itemize}
\end{quote}
\end{center}


\begin{abstract}
  After summarizing basic features of self-organization such as entropy
  export, feedbacks and nonlinear dynamics, we discuss several examples
  in biology. The main part of the paper is devoted to a model of active
  Brownian motion that allows a stochastic description of the active
  motion of biological entities based on energy consumption and
  conversion. This model is applied to the dynamics of swarms with
  external and interaction potentials.  By means of analytical results,
  we can distiguish between translational, rotational and amoebic modes
  of swarm motion. We further investigate swarms of active Brownian
  particles interacting via chemical fields and demonstrate the
  application of this model to phenomena such as biological aggregation
  and trail formation in insects.
\end{abstract}              

\newcommand{\mean}[1]{\left\langle #1 \right\rangle}
\newcommand{\abs}[1]{\left| #1 \right|}
\newcommand{\la}{\langle}
\newcommand{\ra}{\rangle}
\newcommand{\RA}{\Rightarrow}
\newcommand{\tet}{\vartheta}
\newcommand{\eps}{\varepsilon}
\newcommand{\bbox}[1]{\mbox{\boldmath $#1$}}
\newcommand{\ul}[1]{\underline{#1}}
\newcommand{\ol}[1]{\overline{#1}}
\newcommand{\non}{\nonumber \\}
\newcommand{\no}{\nonumber}
\newcommand{\eqn}[1]{eq. (\ref{#1})}
\newcommand{\Eqn}[1]{Eq. (\ref{#1})}
\newcommand{\eqs}[2]{eqs. (\ref{#1}), (\ref{#2})}
\newcommand{\pics}[2]{Figs. \ref{#1}, \ref{#2}}
\newcommand{\pic}[1]{Fig. \ref{#1}}
\newcommand{\sect}[1]{Sect. \ref{#1}}
\newcommand{\name}[1]{{\rm #1}}
\newcommand{\vol}[1]{{\bf #1}}
\newcommand{\et}{{\it et al.}}
\newcommand{\D}{\displaystyle}
\newcommand{\T}{\textstyle}
\newcommand{\SC}{\scriptstyle}
\newcommand{\SSC}{\scriptscriptstyle}
\renewcommand{\textfraction}{0.05}
\renewcommand{\topfraction}{0.95}
\renewcommand{\bottomfraction}{0.95}
\renewcommand{\floatpagefraction}{0.95}
\newcommand{\fn}[1]{\footnote{ #1}}
\vspace*{1cm}
\begin{flushright}
\begin{minipage}{12cm} \small \emph{
  ``Every theory, whether in the physical or biological
        or social sciences, distorts reality in that it 
        oversimplifies. But if it is a good theory, what is 
        omitted is outweighted by the beam of
        light and understanding thrown over the diverse facts.''} \hfill
\textsc{Paul A. Samuelson}
\end{minipage}\hspace*{1.cm}
\end{flushright}

\section{Introduction}
\label{1}

About 1845, \textsc{Hermann von Helmholtz}, the great pioneer in applications of
physics to biological systems, developed the concept \emph{``Physics of
  life''} (\textsc{Markl} 1995) in companionship with his fellows
\textsc{Emil Du Bois-Reymond}, \textsc{Ernst Wilhelm von Br\"ucke} 
and \textsc{Carl Ludwig}.  Their
statement, that life is not in contradiction to physical laws, was later
also elaborated by \textsc{Ludwig Boltzmann} and others. But only in the
20$^{\mathrm{th}}$ century, the investigations of \textsc{Erwin Schr\"odinger},
\textsc{Max Delbr\"uck}, \textsc{Ludwig von Bertalanffy}, 
\textsc{Ilya Prigogine},
\textsc{Manfred Eigen}, \textsc{Mikhail Volkenstein} and others led us to some
understanding of the necessary conditions for the evolution of living
systems (\textsc{Volkenstein} 1994). Their success was based on a
specific theoretical approach to biological problems that also implied
some reductionism. ``Many biologists do not believe that ... biology can
be given a theoretical foundation.  Rather they insist in a holistic
approach. ...  Physicists, on the other hand, have not always appreciated
that a theory of biology has to start from biological facts.  They often
thought that biology is just another field to which they could
immediately apply their equations''. This quotation from \textsc{Eigen's}
foreword to \textsc{Volkenstein's} (1994) book indicates that the road to
a fruitful collaboration between physicists and life scientists -- the
\emph{``Helmholtz road''} -- is full of obstacles.  Nevertheless, we share
the view that at the end of this road we are lead to some useful results,
at least to some better understanding of biological facts. This shall 
also be demonstrated by the examples discussed in the following sections.

We start our considerations with some general remarks on
self-organization and non-linear dynamics in biology. In particular, we
summarize some basic physical principles that lead to the emergence of
complex structures in biological systems, such as openess,
irreversibility, entropy export and feedback processes.  It is well known
from the thermodynamics of irreversible processes that systems may
exhibit a rich variety of complex behavior if there is a supercritical
influx of free energy. This energy may be provided in different forms,
i.e. matter (chemical components, resources), high temperature radiation,
or signals. What kind of complex behavior is observed in a system, will
of course not only depend on the influx of energy but also on the
interaction of the entities that comprise the system.  Among the
prominent examples that can be observed in biological systems are
processes of pattern formation and morphogenesis and different types of
collective motion, such as swarming.

The main part of our paper is devoted to the modelling of active motion
and coherent motion that in biological systems can be found on different
scales, ranging from cells or simple microorganisms up to higher
organisms, such as bird or fish.  Our investigations are based on a model
of active Brownian particles, i.e. particles with an internal energy
depot that can be used for active movement. Considering further
non-linear interactions between the particles, such as attractive forces
or interactions via chemical fields, we are able to derive
a rather general framework for the dynamics of swarms.

By means of both, computer simulations and analytical investigations, we
demonstrate how the superposition of simple microscopic motions may
result in a quite complex dynamics of the macroscopic system.  In
particular, we derive analytical expressions for the distribution
functions that allow to distingush between different modes of swarming
behavior, such as translational, amoebic and rotational modes of
collective motion. Eventually, we study the dynamics of swarms coupled to
chemical fields and demonstrate the application of this model to
phenomena such as biological aggregation and trail formation in insects.

\section{Self-Organization and Nonlinear Dynamics in Biology}
\subsection{General Aspects of Complexity}

>From our daily life experience we know how fragile and complex
biological, ecological and social systems behave. What do we mean by the
term \emph{``complexity''} in a scientific context?  According to our
view complex systems are comprised of multiple components which interact
in a nonlinear manner (cf. Fig. 1), thus the system behavior cannot be
inferred from the behavior of the components. More specifically, these
systems are characterized by (\textsc{Ebeling} \emph{et  al.} 1998):
\begin{itemize}
\item
structures with many components,
\item
dynamics with many modes,
\item
hierarchical level structures,
\item
couplings of many degrees of freedom,
\item
long-range spatial-temporal correlations.
\end{itemize}
\begin{figure}[htbp]
\centering{
\hfill \epsfig{figure=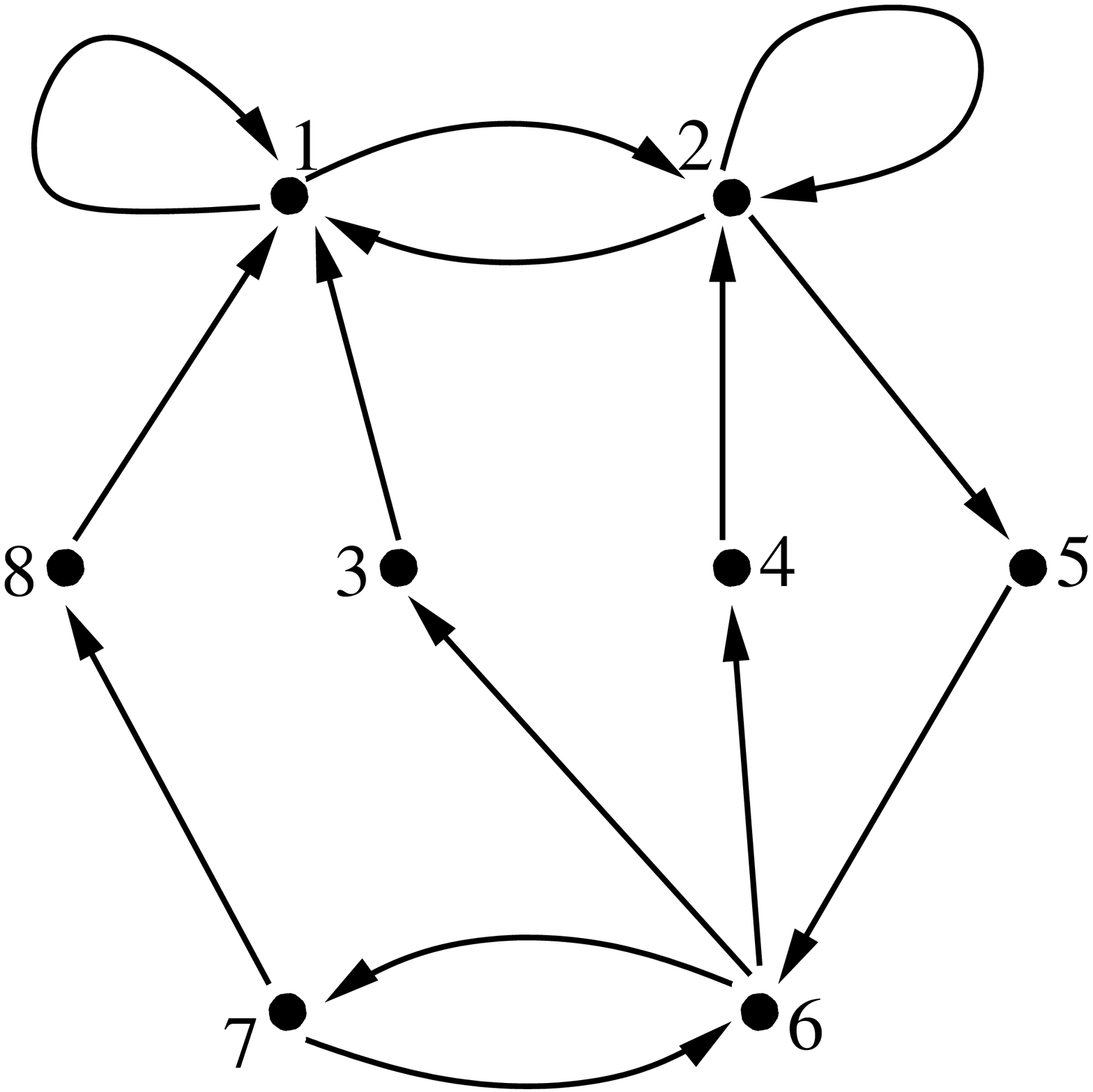,width=5cm}\hfill
\epsfig{figure=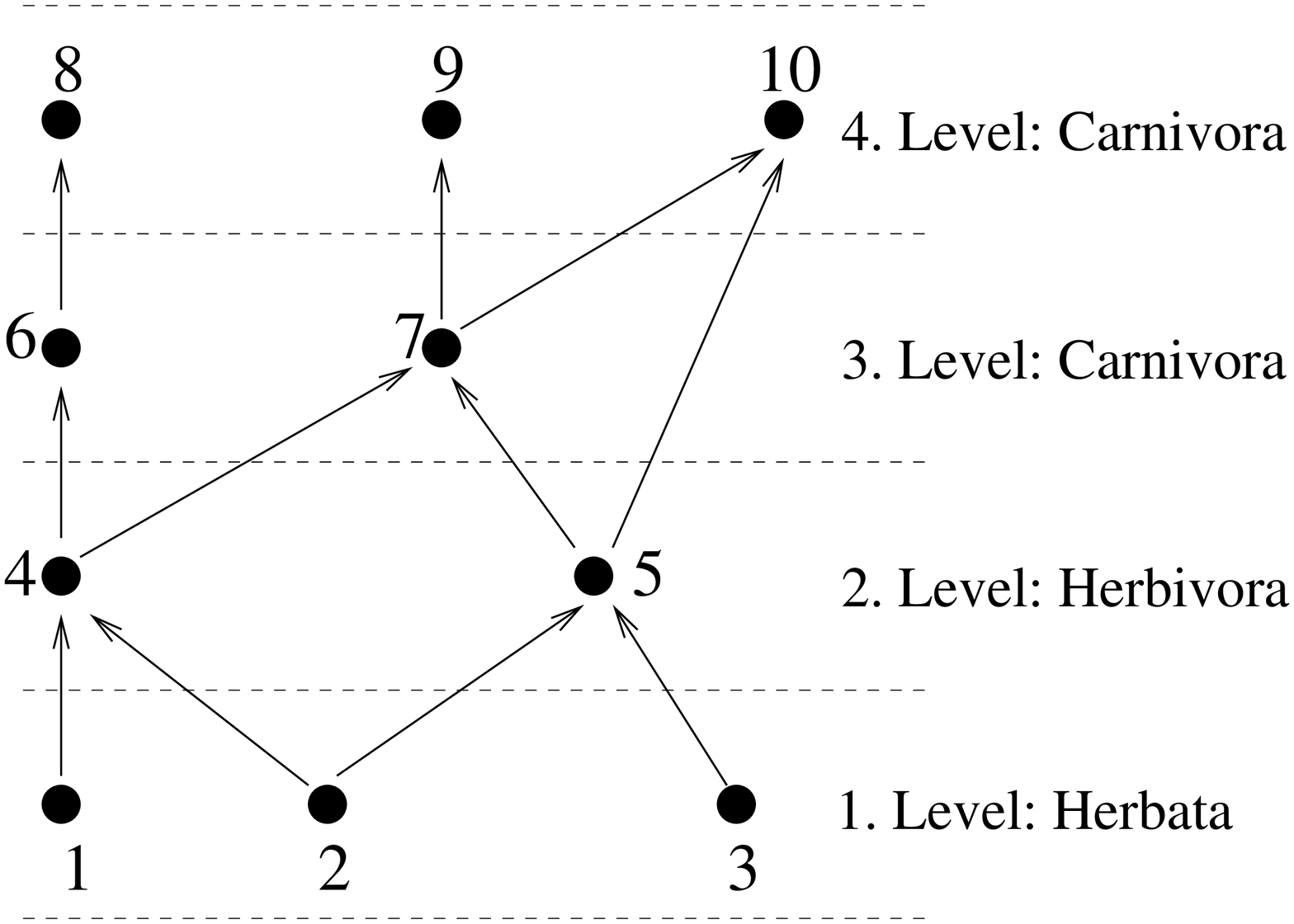,width=7cm}\hfill
}
    \caption{Two graphical representations of the interaction in complex
      systems: (left) a catalytical network consisting of 8 elements with
      14 feedbacks, (right) a hierarchical ecological network. }
    \label{network}
\end{figure}

As we have learned from nonlinear dynamics, complexity is not restricted
to large hierarchical systems, also relatively simple dynamical models
may show complicated behavior. Among the specific features of complex
nonlinear processes, we mention:
\begin{itemize}
\item complicated trajectories and chaos,
\item manifolds of spatial-temporal structures,
\item the limited predictability of future behavior (positive
  Kolmogorov-Sinai entropy).
\end{itemize}
Further, we note that complexity may arise in dissipative as well as in
conservative systems.  In general complex systems in nature and society
are of dissipative nature, i.e. they are based on energy ``consumption''
that allows self-organization processes.  This, however, needs some
physical requirements, such as:
\begin{itemize}
\item thermodynamic openess, i.e. the system exchanges energy, entropy
  and matter with the environment,
\item that on average the system exports entropy, i.e. it imports energy of
  high value and exports energy of low value (cf. Fig. 2),
\item that the system operates far from equilibrium, beyond a critical
  distance from the equilibrium state (cf. Fig. 3),
\item that the causal relations in the system include (positive and negative)
  feedback and feedforward processes (cf. Fig. 1), i.e.  the dynamics of
  the system is nonlinear.
\end{itemize}
For further details we refer to the literature (e.g. \textsc{Ebeling}
\emph{et al.} 1990, \textsc{Ebeling} and \textsc{Feistel} 1994).
\begin{figure}[htbp]
  \centering{\epsfig{figure=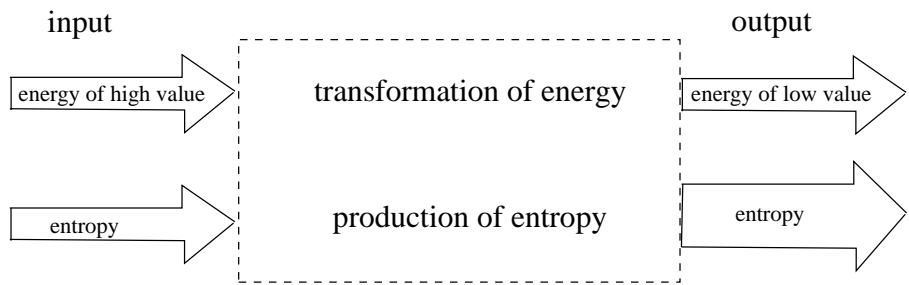,width=12cm}}
    \caption{Transformation  of energy and production of entropy in an
      open system: the export of entropy is a \emph{conditio sine qua
        non} for self-organization.
    \label{entropy}}
\end{figure}
\begin{figure}[htbp]
\centering{\epsfig{figure=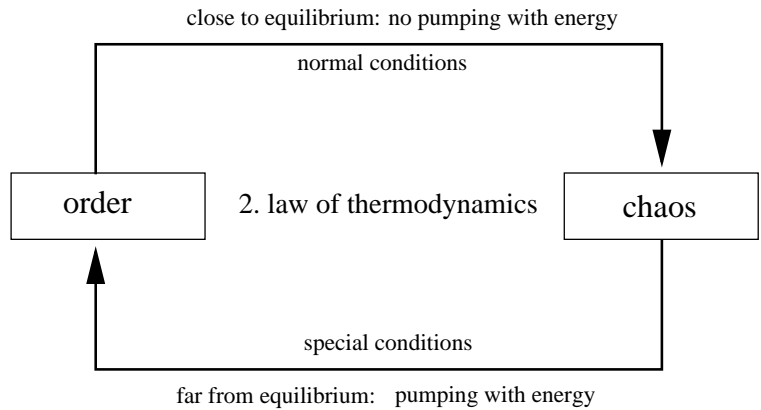,width=10cm}}
    \caption{The 2$^{\mathrm{nd}}$ law of thermodynamics allows
      different processes. Pumping with energy leads the system into
      states far from equilibrium which may be characterized by the
      emergence of ordered structures.  The relaxation into the
      thermodynamic equilibrium, on the other hand, is accompanied by
      the dissappearence of ordered structures.
    \label{order}}
\end{figure}

\subsection{Examples for physical models of biological systems}

It is not intended here to give a complete overview of the vast applications
of physical methods and tools to biological systems. Rather, we \emph{pars
  pro toto} mention here only a few examples, where models based on the
theory of self-organization and non-linear dynamics have contributed to our
understanding of biological phenomena (cf. also the other contributions
in this volume and references therein):
\begin{itemize}
\item \emph{Morphogenesis and biological pattern formation}: After the
  pioneering work on morphogenesis by \textsc{Turing}, \textsc{Meinhard},
  \textsc{Gierer} and others, today a well established theory on
  biological pattern formation exists that is based on the
  reaction-diffusion dynamics of several chemical components
  (morphogens).  It has been successfully applied to a range of
  phenomena, such as patterning of animal coats or sea shells, pattern
  formation in bacterial colonies or slime molds, biological aggregation
  -- but also to processes of regeneration and wound healing, organ
  differentiation, etc.  (see e.g., the contribution by
  \textsc{Holstein}, this volume).

\item \emph{Biological rhythms and synchronization phenomena:} Given that
  the various functional units in biological systems act on different
  time and length scales, the emergence of synchronized behavior is by no
  means self-evident. Recent research in this direction has shown for
  instance how the brain activity is synchronized, or how cardiac cycles
  are triggered by excitation waves.  As another example, the essential
  role of noise could be revealed in the case of stochastic resonance
  (see e.g. the contributions by \textsc{Balaszi},  
  \textsc{Braun}, \textsc{Kantz}, \textsc{Mittag}, \textsc{Moss}, 
  and \textsc{Singer}, this volume)

\item \emph{Directed transport and molecular motors:} The
  ability of living cells to generate motion and forces, e.g. for
  mobility, contraction of muscles or material transport, could recently
  be understood within a physical description. For example, biological
  motor proteins which move along linear filaments can be described by
  stochastic models coupled to chemical reactions.  So-called ratchet
  models further explain the generation of directed motion on the
  microscopic level out of an undirected Brownian motion (see, e.g., the
  contribution by \textsc{H\"anggi}, this volume)

\item \emph{Neural networks and associative memory:} The brain as one of
  the most complex systems known in biology, has also attracted the
  research activities of physicists since the pioneering work of
  \textsc{Hodgkin}, \textsc{Huxley}, \textsc{Hebb}, \textsc{Hopfield} and
  many others. It became clear that information is encoded not only in
  the response of the individual neural cells but also in the joint
  activity of a population of neurons. Based on these investigations, new
  techniques for information storage in associative memories or for
  pattern recognition, but also for brain stimulation have been
  developed.  Artificial neural neural networks today also find a wide
  application in analyzing complex data sets (see e.g. the contributions
  by \textsc{Singer}, by \textsc{Gr\"un} and by \textsc{Tass}, this
  volume)
\end{itemize}

Despite a lot of successful investigations, we have to admit that many
problems in the (physical) understanding of biological processes are still
unsolved. Among the most important is the nature and the origin of
biological information processing (\textsc{Ebeling} and \textsc{Feistel}
1994, \textsc{Volkenstein} 1994).

In the following, we will restrict the discussion to a particular
example, namely \emph{active biological motion}, where we will show in
more detail how a physical approach can be derived and on what reductions
it is based.

\section{Modeling active Brownian movement}
\subsection{Some historical remarks}
Brownian motion denotes the erratic motion of a small, but larger than
molecular, particle in a surrounding medium, e.g. a gas or a liquid.
This erratic motion results from the random impacts between the atoms or
molecules of the medium and the (Brownian) particle, which cause changes
in the direction and the amount of its velocity, $\bbox{v}$.

The motion of the particle is named after the British botanist
\textsc{Robert Brown} (1773-1858), who in 1827 discovered the erratic
motion of small pollen grains immersed in a liquid.  He was inclined to
explain his observation by so-called ``active molecules'', and it is also
reported that he wrote a letter to \textsc{Charles Darwin} to ask him about his
opinion on this subject.

\textsc{Brown} was not the first who observed such a motion with a
microscope. For example, already the Dutch \textsc{Anton van Leeuwenhoek}
(1632-1723), who first discovered micro organisms with a simple
microscope, knew about the typical erratic motion, however he considered
it a feature of living entities. In 1785, the Dutch physician \textsc{Jan
  Ingenhousz} (1730-1799) also reported the erratic motion of anorganic
material dispersed in a liquid, i.e. powdered charcoal floating on
alcohol surfaces, but this became not known to the non-Dutch speaking
world.

The physical explanation of Brownian motion started about 1900 with the
seminal works of \textsc{Albert Einstein} (``On the theory of Brownian
motion'' 1905) and \textsc{Marian Smoluchowski} (``On the kinetic theory
of Brownian molecular motion and suspensions'' 1906), but it should be
noticed that already in 1900 \textsc{Louis Bachelier} has derived a
mathematical theory of this type of stochastic processes while
investigating price changes at the stock market.

Brownian motion would be rather considered as \emph{passive motion},
simply because the Brownian particle does not play an active part in this
motion. It is an \emph{undirected} motion, driven by thermal noise.
Passive motion can also be directed, e.g. if it is driven by convection,
currents or by external fields.  Active motion, on the other hand relies
on the supply of energy, i.e. it occurs under energy consumption and
energy conversion and may also involve processes of energy storage. In
order to add such a new element to the concept of Brownian motion, we
need to investigate possible mechanisms of energy pumping.

The idea of energy supply was first introduced in the context of the
theory of sound and music by \textsc{Helmholtz} (``Die Lehre von den
Tonempfindungen'' 1870) and \textsc{Rayleigh} (``On maintained
vibrations'' 1883; ``The theory of sound'' 1877).  The \textsc{Rayleigh}
model of self-sustained oscillations is based on the assumption of a
velocity-dependent friction coefficient (cf. Fig. 4), that can be
negative for a certain range of velocity, i.e. instead of dissipating
energy because of friction, energy can be pumped into the system.  That
means if a violin bow transfers energy to the string via friction
\emph{negative friction} occurs. Provided a supercritical influx of
energy, a self-sustained periodic motion can be obtained, namely the
violin string emits acoustic waves.
\begin{figure}[htbp]
  \centerline{\psfig{figure=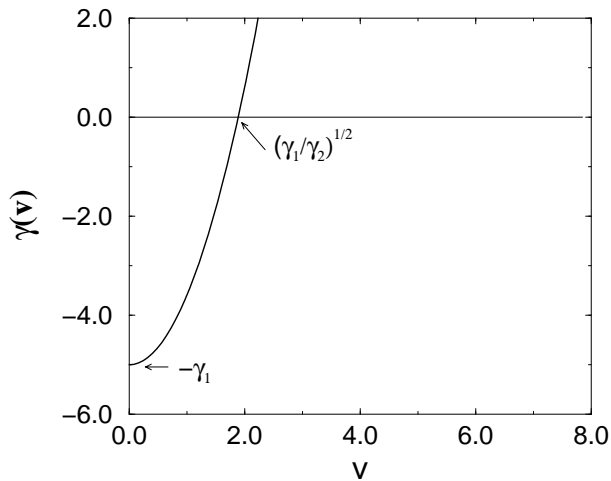,width=8cm}}
  \caption[]{\label{gamma-ray-grul}
    \textsc{Rayleigh}-type velocity-dependent friction coefficient
    $\gamma(\bbox{v}) = - \gamma_1 + \gamma_2\, \bbox{v}^2$.  For
    $\gamma(\bbox{v})<0$ ``pumping'' dominates, while for
    $\gamma(\bbox{v})>0$ ``dissipation'' dominates.}
\end{figure}

In the following we will show how the integration of the ideas about
maintained vibrations and Brownian motion leads to a new model of
\emph{active movement}. This term will be used now for all kinds of motions
in space and time which are driven by sources of free energy.

\subsection{Brownian particles with energy supply}

In this section, we introduce a simple stochastic model of active
movement called ``model of active Brownian particles''
(\textsc{Schweitzer} \emph{et  al.} 1998, \textsc{Ebeling} \emph{et al.}
1999). Let us consider $i=1,...,N$ active Brownian particles 
with mass $m$ located at
the positions $\bbox{r}_{i}$ and moving with the velocity $\bbox{v}_{i}$.
For the equation of motion we postulate:
\begin{equation}
\label{motion}
m \frac{d \bbox v_i}{d t} + \frac{\partial U}{\partial \bbox r_i} =
\bbox F_i   + \sqrt{2 D}\, \bbox \xi_i (t)
\end{equation}
The last term denotes the stochastic force acting on the Brownian
particle $i$ with a strength $D$, the random function $\bbox{\xi}_{i}(t)$
is assumed to be Gaussian white noise.  $U$ is the potential of external
and interaction forces and $\bbox F_i$ is the dissipative force acting on
particle $i$. It can be specified as:
\begin{equation}
\label{fi}
\bbox F_i = -m \gamma \bbox v_i + d e_i \bbox v_i
\end{equation}
Here $\gamma$ is the usual passive friction coefficient with the
dimension of a frequency.  We assume that the noise intensity $D$ is
related to the friction by an Einstein relation $D = m \gamma kT$,
where $k$ is the \textsc{Boltzmann} constant and $T$ is the temperature.  
The
second term $(d \bbox v_i e_i)$ expresses an acceleration of the particle
in the direction of $\bbox v_i$ (a forward thrust) which is based on the
conversion of energy from a internal energy depot $e_{i}$ of the
particle.  More specifically, we assume that the Brownian particle is able to
take up energy with the rate $q$, which can be stored in an internal
depot $e_{i}$.  The internal energy can be converted into kinetic energy
with a momentum dependent rate $m d v_i^2$, which results in the
acceleration in the direction of movement.  The internal energy
dissipates with the rate $c e_{i}$, The balance is then expressed by:
\begin{equation}
\label{balance}
\frac{d e_{i}}{d t} = q - c e_{i} - d v_{i}^2 e_{i}
\end{equation}
If the internal energy depot relaxes fast compared to the motion of the
particle, we find for \eqn{fi} in adiabatic approximation:
\begin{equation}
\label{adiabat}
\bbox F_i = - m \bbox v_i \, g(v_i^2) = \bbox v_i \,
\left(\frac{d q}{c + d v_i^2} - m \gamma \right)
\end{equation}
Here $g(v^{2})$ denotes a velocity-dependent \emph{friction function}.
>From now on we will use units corresponding to $m=1$, i.e.  $\bbox v = \bbox
p$.  Dependent on the parameter values, the dissipative force
$\bbox{F}_{i}$ may have one zero at $\bbox v = 0$, or two more zeros with
\begin{equation}
  \label{v-0}
  \bbox{v}_0^2=\frac{q}{\gamma} - \frac{c}{d}
\end{equation}
A nontrivial velocity $\abs{v_{0}}>0$ only exists if $qd > c \gamma$,
i.e.  if a supercritical supply of energy occurs. In this case, we also
speak about \emph{``active particles''}.  For $0 < \abs{v} <
\abs{v_{0}}$, i.e.  in the range of small velocities the dissipative
force $\bbox{F}_{i}$ is positive, i.e.  the particle is provided with
additional free energy.  On the other hand, for $0 < \abs{v_{0}} <
\abs{v}$, the dissipative force is negative. Hence, slow particles are
accelerated, and fast particles are decelerated.

Assuming $v_0^{2} > 0$ we consider now two limiting cases.  Introducing
the bifurcation parameter $\zeta = (d q / c \gamma) -1$, we get for
small values of the parameter $\zeta$ the well-known law of Rayleigh (cf.
Fig. 4):
\begin{equation}
\label{f-rayleigh}
\bbox F = \gamma \zeta \left(1 - \frac{v^2}{v_0^2}\right) \bbox v
\end{equation}
In the opposite case, i.e. for large values of $\zeta$, we get the
empirical law found by \textsc{Schienbein} and \textsc{Gruler} (1993) for
the dynamics of cells:
\begin{equation}
\label{f-gruler}
\bbox F = 2 \gamma \left( 1 - \frac{v_0}{v}\right) \bbox v
\end{equation}
This way, our expression for the dissipative force $\bbox{F}_{i}$,
\eqn{adiabat} is general enough to cover interesting limiting cases. We
mention that in other models of driven motion (\textsc{Viscek} \emph{et
  al.} 1995) the velocity $v_{0}$ is postulated without further
investigations.

\subsection{Velocity distribution and mean squared displacement of free
  active motion}

We are now interested in how known features of Brownian motion, such as the
stationary velocity distribution or the mean squared displacement, change
if we consider a supercritical energy take up ($qd>c\gamma$) of the
Browian particles.
In order to find the velocity distribution explicitely we have to
formulate and to solve the Fokker-Planck equation corresponding to Eq.
(1).  We restrict our consideration here to the two-dimensional space and
$U=0$, i.e. there are no external or interaction forces.  Following standard
procedures (\textsc{Klimontovich} 1995), we find from the Fokker-Planck
equation the stationary solution for the velocity distribution
(\textsc{Erdmann} \emph{et al.} 2000):
\begin{equation}
  \label{f0-v}
  P^{0}(\bbox v)= C\,\left(1+ \frac{d}{c} \bbox v^2 \right)^{\frac{q}{2D}}\;
  \exp{\left( - \frac{\bbox v^2}{2 k T} \right)}
\end{equation}
Compared to the Max\-well\-ian velocity distribution of ``simple''
Brownian particles, a new prefactor appears in Eq.(8)
which results
from the internal energy depot. For supercritical pumping, $q d > \gamma
c$, we find a crater-like velocity distribution, which indicates strong
deviations from the Maxwell distribution (cf. Fig. 5).
\begin{figure}[htbp]
  \centerline{\psfig{figure=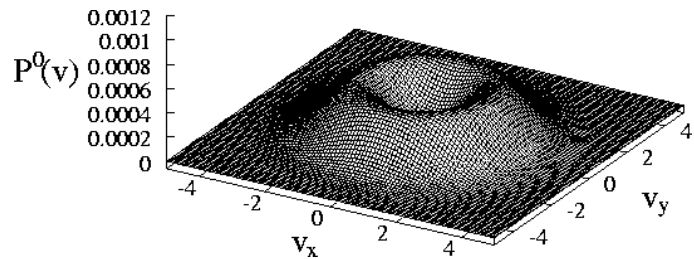,width=9cm}}
  \caption[]{\label{v-dist}
   Stationary velocity distribution $ P^{0}(\bbox v)$ for active
   Brownian particles in the case of supercritcal
   energy supply (\textsc{Erdmann} \emph{et al.} 2000). }
\end{figure}

The distribution represnted by Eq.(8)
is an exact result for non-interacting
particles.  In the limit of zero noise, $D \rightarrow 0$, it obtains the form
$\delta(v^2 - v_0^2)$.  In this small noise limit, a  result for the
\emph{mean square displacement} is also available (\textsc{Erdmann}
\emph{et al.} 2000):
\begin{equation}
\label{mean-square}
\mean{\left(\bbox{r}(t) - \bbox{r}(0)\right)^2} = \frac{2 v^4_0}{D} t +
\frac{v_0^6}{D^2}\left[\exp{\left(-\frac{2Dt}{v_0^2}\right)}-1\right]
\end{equation}
>From Eq.(9),
we find the effective spatial diffusion
coefficient of active Brownian particles as $D_{r}^{\mathrm{eff}}=v_0^4 /
D$. This expression leads to rather large values for small $D$ or large
$v_0$.  The analytical expressions for the stationary velocity
distribution and for the mean square displacement are in good agreement
with computer simulations (\textsc{Schweitzer} \emph{et al.} 2001) and
with measurements on the active movements of granulozytes
(\textsc{Schienbein} and \textsc{Gruler} 1993). We suggest to compare
them also with the observations of the movement of \emph{Daphnia} (see
\textsc{Ordemann} and \textsc{Moss}, this volume).

\section{Swarm dynamics with external and interaction potentials}

\subsection{Dynamics in external potentials}

Let us now consider a swarm of active particles in a two-dimensional
radially symmetric potential $U(r)= a (x_1^2+x_2^2)$, that generates an
attractive force towards the center, $r = 0$.  As the snapshot of the
spatial dispersion of the swarm shows Fig.6, we find 
the occurence of two branches of the swarm, which after a
sufficiently long time move on \emph{two limit cycles}.  One of these
limit cycles refers to the left-handed, the other one to the right-handed
direction of motion in the 2d-space.
\begin{figure}[htbp]
\centerline{\psfig{figure=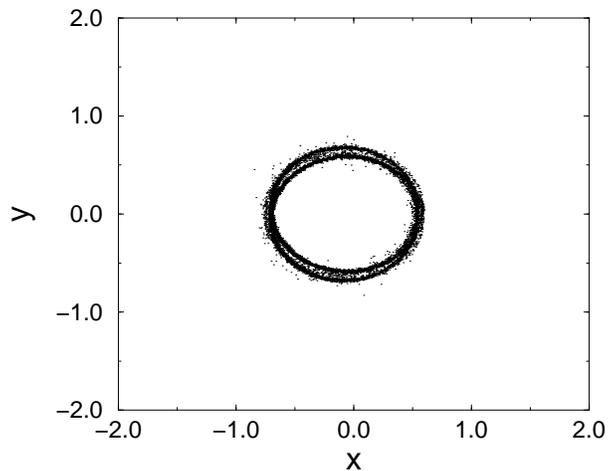,width=8.cm}}
\caption[]{\label{fs-swarm}
  Spatial snapshot at $t=99$ of a swarm of $10.000$ active particles
  moving in a parabolic potential (\textsc{Schweitzer} \emph{et al.}
  2001).}
\end{figure}

The radius of the limit cycles can be calculated with the following
considerations: Moving under stationary conditions, the particles have to
comply with the additional requirement to balance between centrifugal and
attracting forces, which leads to the condition $v^2/r = |U'(r)|$.  For
the harmonic potential this results in the stationary radius $r_0 = v_0 /
\omega_0$ where the frequency of rotations is given by $\omega_0^2 = a/m$
(\textsc{Erdmann} \emph{et al.} 2000).

For the motion on the limit cycle, an exact solution of the equations of
motion reads in the deterministic limit:
\begin{equation}
  \label{4d-explizit}
  \begin{aligned}
    x_1 = r_0\, \cos{(\omega_{0} t + \phi_0)} & \qquad
    v_1 = - r_0\, \omega_{0}\, \sin{(\omega_{0} t + \phi_0)}\\
    x_2 = r_0\, \sin{(\omega_{0} t + \phi_0)} &\qquad
    v_2 = r_0\, \omega_{0}\, \cos{(\omega_{0} t + \phi_0)}
  \end{aligned}
\end{equation}
Another exact solution is obtained by inversion. Any initial state
converges to one of these attractor states. In the presence of
fluctuations, the particles move in the neighbourhood of these two limit
cycle orbits,  which have circle-like projections and are located on two
planes corresponding to the angular momenta $L = \pm v_0^2 / \omega$.
In this way, the probability is concentrated on two toroids in the
4-dimensional phase, the stationary distribution may be approximated by:
\begin{equation}
\label{full}
  \begin{aligned}
    P^{0} (x_1, x_2, v_1, v_2) = &\,C \left[1 +
      \frac{d}{2c}\left(v^2+\omega^2 r^2\right) \right]^{\frac{q}{2D}}
    \exp{\left[ - \frac{1}{2k_{B}T}
        \left(v^2 + \omega^2 r^2\right) \right]} \times \\
    & \times \left[1 + \frac{d}{2 c}
      \frac{|L|}{r_0^2}\right]^{\frac{q}{2Dr_0^2}}
    \exp{\left[-\frac{1}{2k_{B}T}\frac{L^2}{r_0^2}\right]}
  \end{aligned}
\end{equation}
Here the first factor represents a shell with given energy in the
4-dimensional phase space, while the second factor projects out two
planes perpendicular to the two possible directions of the angular
momentum, $L$. In this way two toroids in the 4-dimensional phase space are
generated where the occupation density is concentrated.

\subsection{Harmonic swarms}

So far we have neglected any coupling within the swarm of active
particles. If the swarm is not bound by an external potential as
discussed above, the absense of interactions leads to the effect that the
swarm eventually disperses in the course of time, whereas a ``real''
swarm would maintain its coherent motion. A common way to introduce
correlations between the moving particles in physical swarm models is the
coupling to a mean value. For example \textsc{Czirok} \emph{et al.}
(1996) discuss the coupling of
the particles' individual \emph{orientation} (i.e. direction of motion)
to the mean orientation of the swarm.  Other versions assume the coupling
of the particles' velocity to a \emph{local average velocity}, which is
calculated over a space interval around the particle (\textsc{Czirok}
\emph{et al.} 1999).

Instead of an external potential $U(\bbox{r})$, let us now assume an
\emph{interaction potential}. As the most simple case we may discuss the
global coupling of the swarm to the \emph{center of mass}.  That means
the particle's position $\bbox{r}_{i}$ is related to the mean position of
the swarm $\bbox{R}=1/N\sum \bbox{r}_{i}$ via a potential
$U(\bbox{r}_{i},\bbox{R})$. For simplicity, we may assume a parabolic
potential, i.e. the Hamiltonian for each particle reads now:
\begin{equation}
\label{hamilton}
H_i = \frac{\bbox v_i^2}{2} + \frac{a}{N} \sum_{j\neq i}
(\bbox r_i -\bbox r_j)^2
\end{equation}
With respect to the harmonic interaction potential we
call such a swarm a \emph{harmonic} swarm (\textsc{Ebeling} and
\textsc{Schweitzer} 2001).  The coupling to the center of mass
corresponds to the assumption that there is now an attractive force
between each two particles $i$ and $j$ which depends linearly on the
distance between them. This can be used to control the dispersion of the
swarm. A special case of nonlinear (exponential) interactions between
particles on a chain has been analyzed in detail by \textsc{Ebeling}
\emph{et al.} (2000).

Fig. 7 presents snapshots of a computer simulation of a harmonic swarm of
2.000 active particles.%
\fn{A movie of these computer simulations -- with the same
parameters, but a different random seed -- can be found at
\url{http://ais.gmd.de/~frank/swarm-tb.html}}%
Due to a supercritcal take-up of energy, the particles are able to move
actively, the interaction however prevents the swarm from simply
dispersing in space. Thus, the collective motion of the swarm becomes
rather complex, as a compromise between \emph{spatial dispersion} (driven
by the energy pumping) and \emph{spatial concentration} (driven by the
mutual interaction).
\begin{figure}[htbp]
\centerline{
\hfill
\psfig{figure=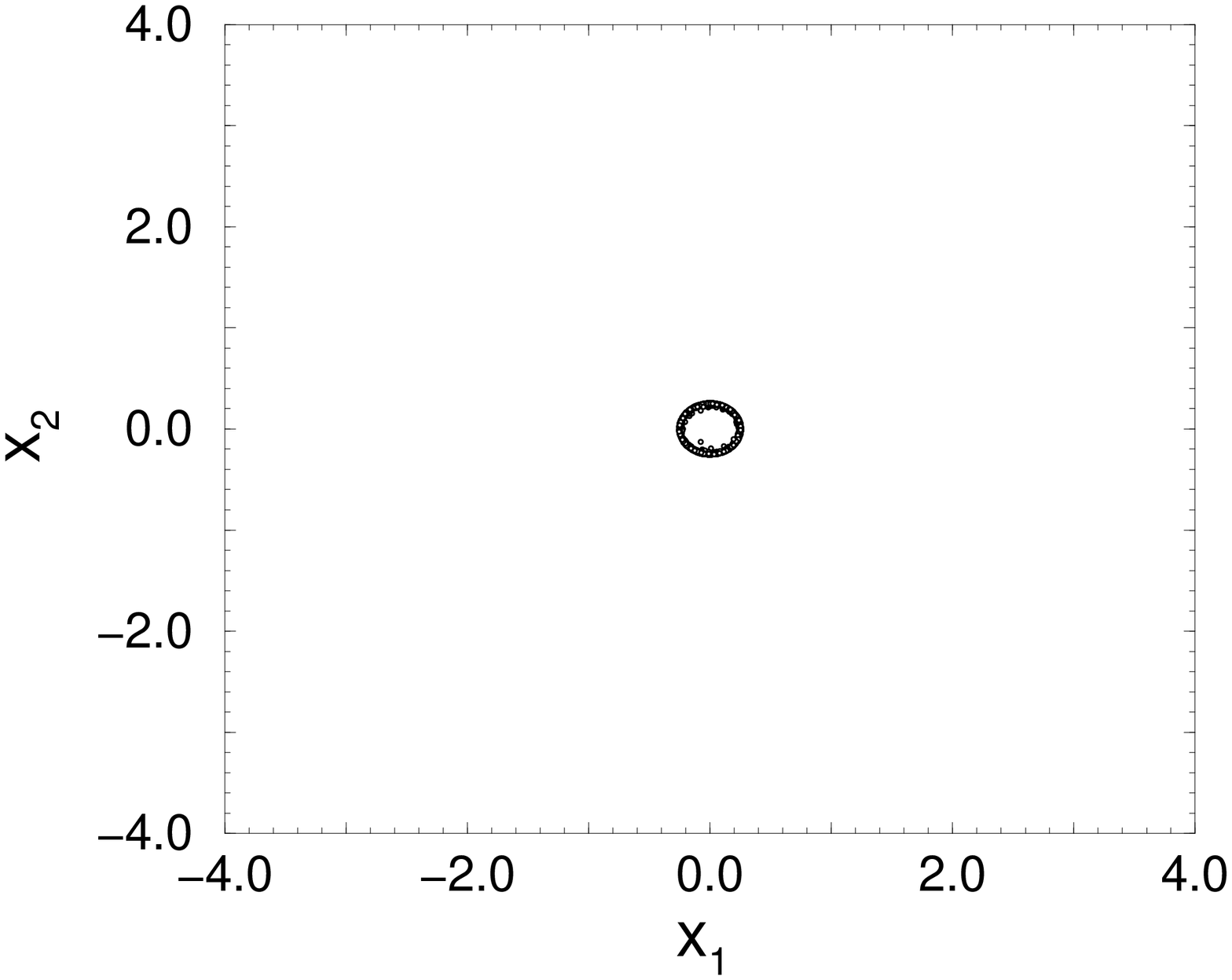,width=5.5cm}
\hfill
\psfig{figure=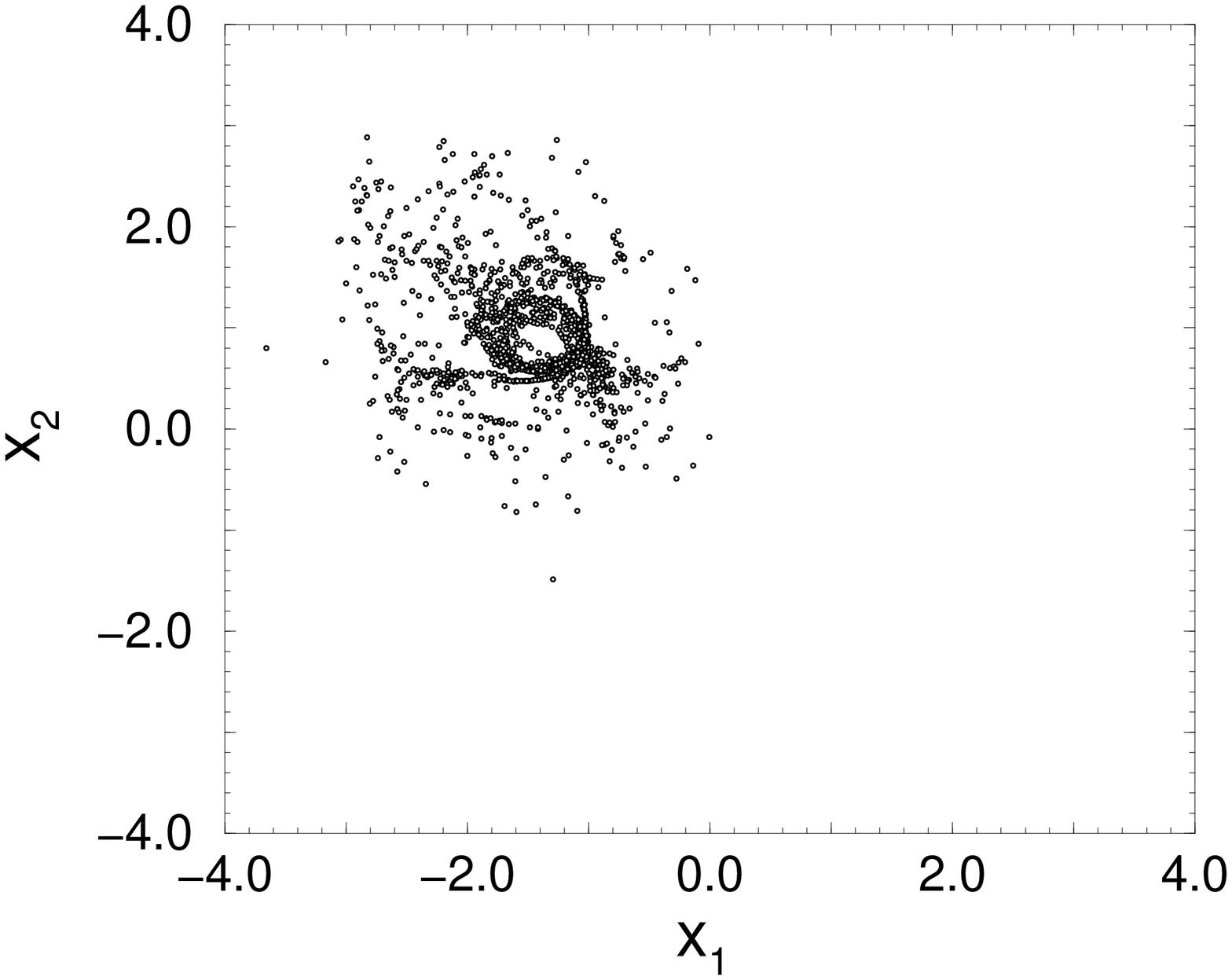,width=5.5cm}
\hfill}
\centerline{\hspace*{3.5cm} t=1 \hfill t=25 \hspace*{3.5cm}}\vspace*{4mm}
\centerline{
\hfill
\psfig{figure=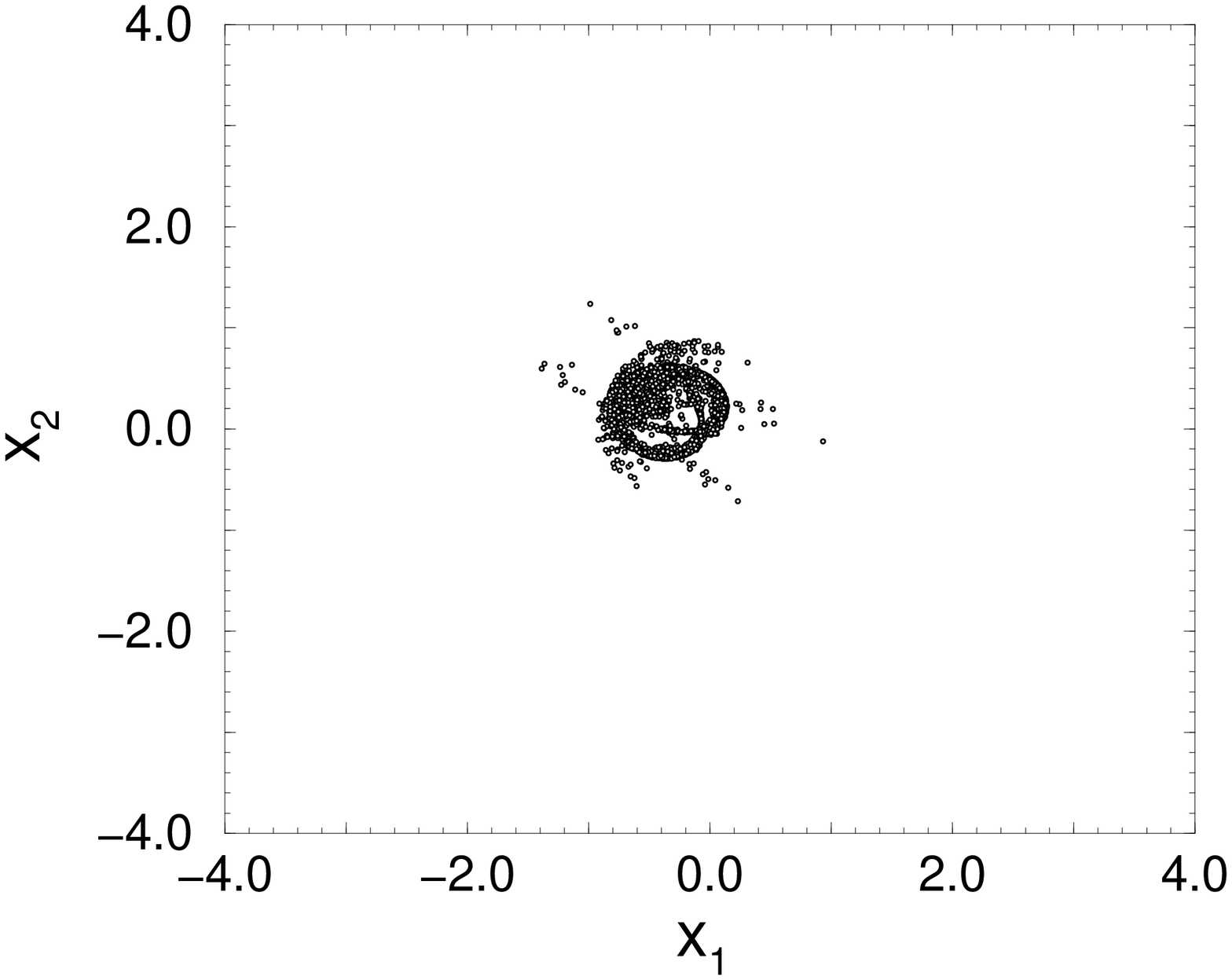,width=5.5cm}
\hfill
\psfig{figure=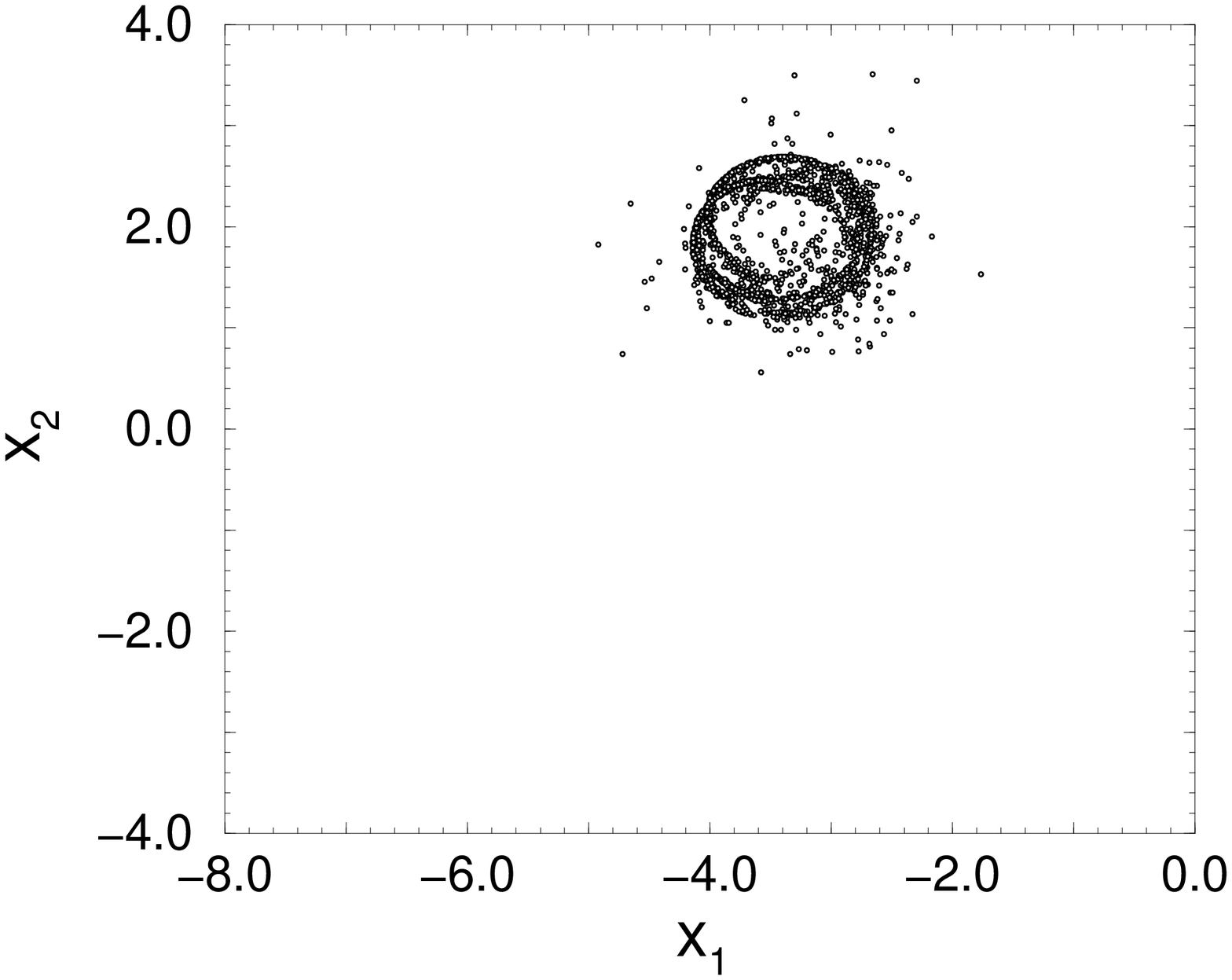,width=5.5cm}
\hfill}
\centerline{\hspace*{3.5cm} t=10 \hfill t=50 \hspace*{3.5cm}}
\caption[]{\label{snap}
  Snapshots (spatial coordinates) of a harmonic swarm of $2.000$ active
  particles. $t$ gives the different times.  Initially, the particles were
  at rest and at the same spatial position.  Note that the picture for
  $t=50$ has a shifted $x_{1}$-axis (\textsc{Ebeling} and
  \textsc{Schweitzer} 2001).}
\end{figure}

A closer inspection of the swarm dynamics (\textsc{Ebeling} and
\textsc{Schweitzer} 2001, \textsc{Schweitzer} \emph{et al.} 2001) reveals
that the system basically possesses two nontrivial dynamic modes.  The
first mode corresponds to a \emph{flock-like swarm} moving coherently
with given direction (translational mode). The second mode corresponds to
a \emph{rotating swarm} while the center of mass is at rest (rotational
mode). Which of these modes is the target (attactor) of the collective
motion depends both on the initial conditions and on the strength of
noise.

Let us now characterize the two modes by means of the distribution
functions.  In the first mode, the particles move parallel to the
velocity of the the center of mass, $V$. Introducing the
relative velocity $\delta \bbox{v}_i = \bbox{v}_i - \bbox{V}$, we get in
first approximation the distribution:
\begin{equation}
\label{p0-trans}
\begin{aligned}
  P^{0} (\bbox r_i, \bbox v_i) = &\,C\left[1 + \frac{d}{c} V^2
  \right]^{\frac{q}{2D}}
  \exp{\left[- \frac{V^2}{2kT}-\frac{a}{2 kT}
(\bbox r_i - \bbox R)^2\right]} \times \\
&\times \exp{\left[-\frac{1}{2\gamma kT} \Big(g(V^2) (\delta
        \bbox v_i)^2 + 2 g'(V^2) (\bbox V \cdot \delta \bbox v_i)^2
      \Big)\right]}
\end{aligned}
\end{equation}
Here, $g(V^{2})$ denotes the friction function introduced in
Eq.(4),
whereas $g'(V^2)$ is the first derivative of $g(V^{2})$.
According to Eq.(13),
the square of the translational velocity
$V^2$ is near to $v_0^2$ and the deviations fluctuate according to the
Boltzmann distribution.

As we have shown by means of computer simulations (\textsc{Schweitzer}
\emph{et al.} 2001), the translational mode breaks down for small initial
velocities $\bbox V^2 \ll v_0 ^2$ (cf. also Fig. 7). 
In this case the velocities
relative to the center of mass are amplified. On the other hand, the
translational mode also becomes unstable if the dispersion of the
relative velocities approaches the order of $v_0^2$. In this way, the
overall picture is similar to the findings for the one-dimensional case
(\textsc{Mikhailov} and \textsc{Zanette} 1999).

In the second stationary mode, where the center of mass is at rest $\bbox
R$ = const., $\bbox V$ = 0, the swarm is rotating around the center of
mass, and we find  in first approximation the distribution, Eq.(11)
again, with $H_{i}$ given by Eq.(12):
\begin{equation}
\label{p0-rot}
  \begin{aligned}
    P^{0} (\bbox{r}_{i}, \bbox{v}_{i}) = &\,C \left[1 +
\frac{d}{2c} H_{i}\right]^{\frac{q}{2D}}
    \exp{\left[ - \frac{H_{i}}{2k_{B}T}
\right]} \times \\
    & \times \left[1 + \frac{d}{2 c} \frac{|L_{i}|}{r_0^2}\right]
^{\frac{q}{2D r_0^2}}
\exp
    \left[-\frac{1}{2k_{B}T}\frac{L_{i}^2}{r_0^2}\right]
  \end{aligned}
\end{equation}
The two possible branches of the rotating swarm correspond to the
positive/negative angular momenta $L = m (x_{1} v_2 - x_{2} v_1)$.

There is still a third mode which is realized in the case of very strong
noise, $k_{B}T \gg m v_0^2$. In this case the system does not feel the
driving force anymore, hence it forms a Boltzmann distributed cluster
with a stochastically moving center:
\begin{equation}
  \label{p0-vB}
  P^{0} (\bbox r_i, \bbox v_i) = C\,\exp{\left[ - \frac{H_i}{k_{B}T}
  \right]}
\end{equation}
In this way, we have -- for a rather special model with linear attraction to
the center -- obtained a full stochastic description of three swarming
modes. Despite our reductionistic approach, our findings agree also with
the qualitative description of \textsc{Okubo} and \textsc{Levin} (2001),
who distinguish between three types of collective animal movement:
\begin{itemize}
\item \emph{Rectilinear movement}: The animals as a whole tend to perform
  a rectilinear movement, thus forming a tight (cohesive) group.
\item \emph{Doughnut pattern}: When the forward thrust dominates the
  random force, a group of animals rotates around an empty center,
  forming the shape of a doughnut.
\item \emph{Amoebic movement}: When the random force dominates the
  forward thrust, the center of mass of animals hardly moves, though the
  shape of the group fluctuates around a circular pattern.
\end{itemize}
Hence, we conclude that even in the rather abstract description of
physical swarm models, basic features of collective motion and swarm
behavior can be recovered and, hopefully, also compared with biological
observations of translating/rotating swarms of fish and birds.

\section{Swarm dynamics in the presence of chemical fields}
\subsection{Models of biological aggregation}

So far, we have assumed in our model that the linear attraction between
any two members of the swarm is of physical nature. The results remain
also valid if there is a \emph{chemical} attraction directed to the
center of mass of the swarm.  This is a reasonable assumption e.g. for
the description of the dynamics of bacterial colonies (\textsc{Vicsek}
2001).  Here, the chemotactic attraction might be responsable for the
widely observed rotational movements of bacteria as \emph{Bacillus
  circulans}, \emph{Clostridium Tetani}, \emph{Paenibacillus vortex}.  If
$A$ is the chemotactic coefficient, the attraction of the active
particles to the center is now given by a linear chemotactic force ${\bbox
  F}_{ch} = - A {\bbox r}$. In this case, the two characteristic
quantities of our distribution functions derived above read  as
$\omega_0^2 = A/m$ and $r_0^2 = v_0^2 m/A$, and the dynamics discussed
above remains the same.

A more elaborated investigation has to consider not only the response of
the particles to the chemical signal, but also the generation of these
chemicals by the particles, i.e. a non-linear feedback between particles
and chemical. In order to describe the chemotactic response of the
particles, we modify the Langevin Eq.(1),
by replacing the potential
$U$ with a scalar field $h(\bbox{r},t)$ that describes the
spatio-temporal concentration of the chemical.  Assuming that the
particles are attracted by higher concentration of the field, we find:
\begin{equation}
\label{motion-h}
m \frac{d \bbox v_i}{d t} + \left.\frac{\partial h(\bbox{r,t})}{\partial \bbox r}\right|_{\bbox{r}_{i}} =
\bbox F_i   + \sqrt{2 D}\, \bbox \xi_i (t)
\end{equation}
In a biological context, the chemical field can for example represent
pheromones produced e.g. by ants or other insects in order to communicate
with their mates, i.e. it can be envisioned as a communication medium that
contains spatial information produced by the insects. The chemotactic
response to the field is a basic feature of phenomena such as trail
formation in ants (\textsc{Edelstein-Keshet} \emph{et al.} 1995,
\textsc{Schweitzer} \emph{et al.} 1997), it also plays an important role
in the formation of biological patterns in bacteria \emph{Escherichia coli}
(\textsc{Ben-Jacob} \emph{et al.} 1994) or slime molds (\textsc{H\"ofer}
1999).

For the dynamics of the chemical field $h(\bbox{r},t)$, we assume the
following reaction-diffusion equation:
\begin{equation}
\frac{\partial h(\bbox{r},t)}{\partial t} =
\sum_{i=1}^{N} s\,
\delta\Big(\bbox{r}-\bbox{r}_{i}(t)\Big)
-k\,
h(\bbox{r},t)
\, +\, D_h \,\Delta_{h} h(\bbox{r},t)
\label{h-net}
\end{equation}
It means that changes of the chemical concentration in space and time are
governed by three processes: (i) production of chemical signals by the
particles with a rate $s$ at their current position, $\bbox{r}_{i}$, (ii)
decay of the chemical with a rate $k$, and (iii) diffusion (coefficient
$D_{h}$).

The nonlinear feedback between the particles and the chemical field
eventually results in the formation of aggregates, as the snapshots in
Fig. 8 show.  Biological aggregation based on chemotaxis is widely found
in biological species, such as insect larvae (\textsc{Deneubourg}
\emph{\et al.} 1990) or myxobacteria (\textsc{Stevens} and
\textsc{Schweitzer} 1997, \textsc{Deutsch} 1999) that gather guided by
chemical signals originated by the individuals.

\begin{figure}[htbp]
\setlength{\fboxsep}{0pt}
\centerline{\hfill
\fbox{\psfig{figure=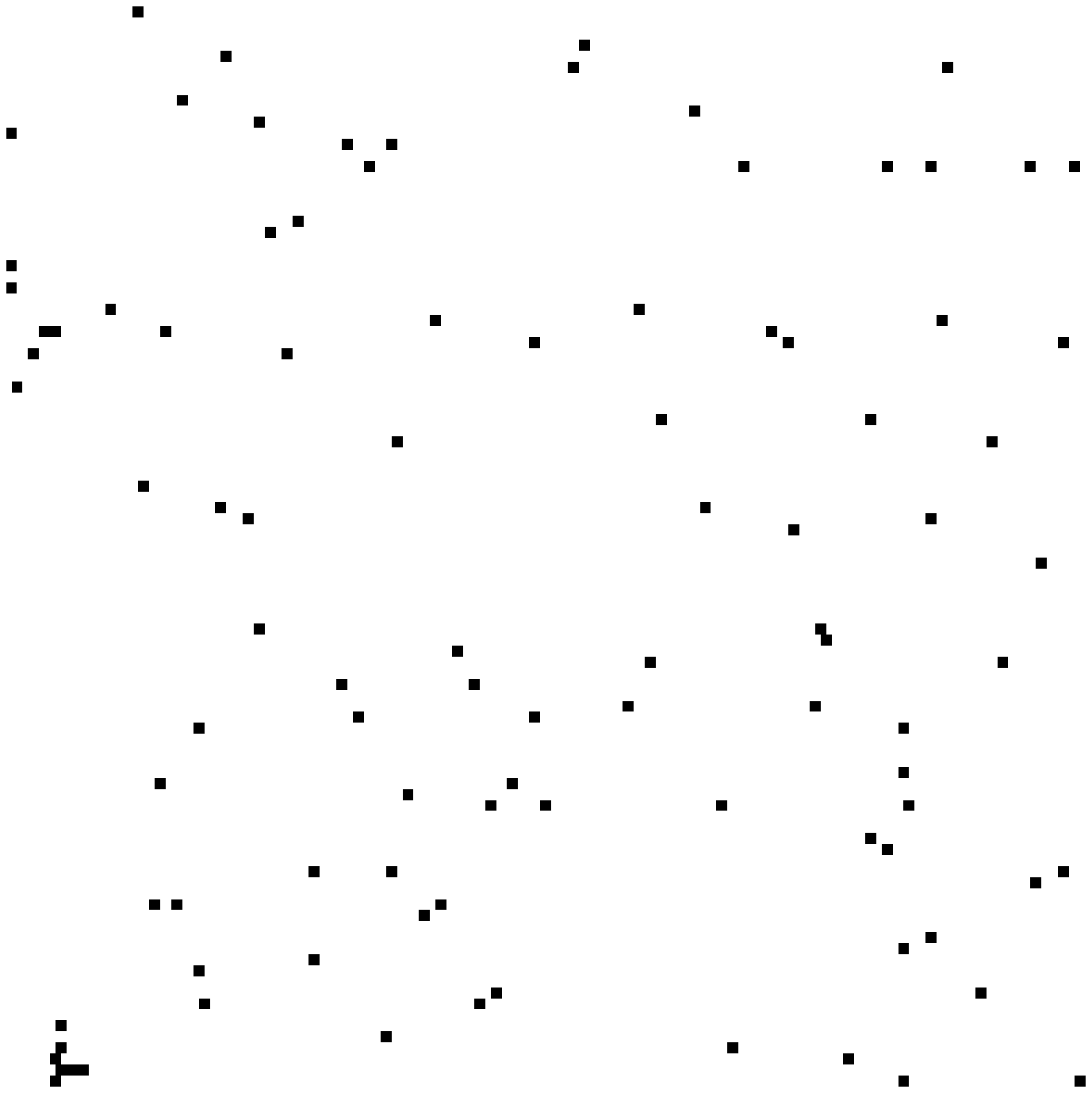,height=4.5cm}}\hfill (a) \hfill
\fbox{\psfig{figure=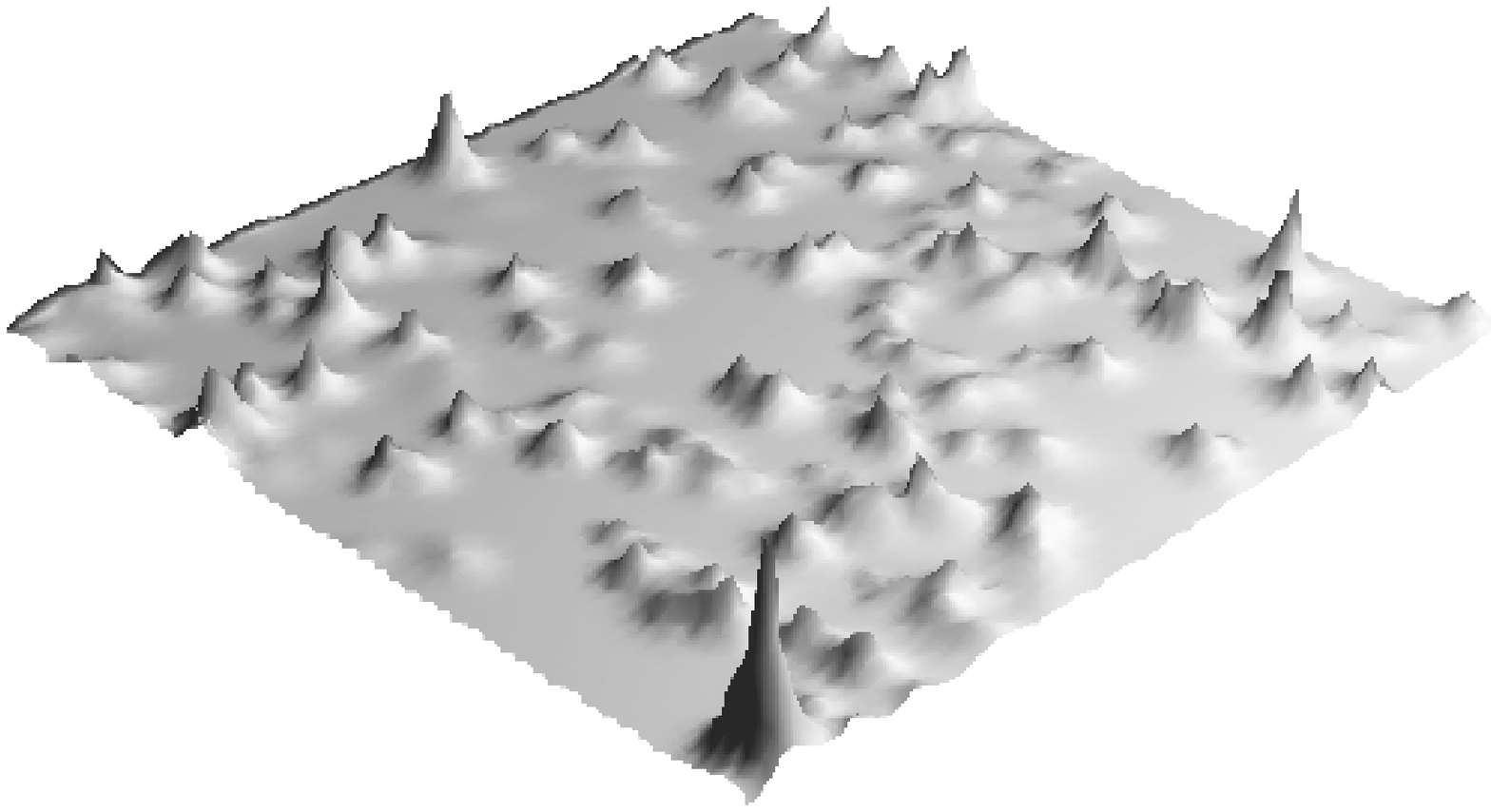,height=4.5cm}}}\hfill
\vspace*{0.3cm}

\centerline{\hfill
\fbox{\epsfig{figure=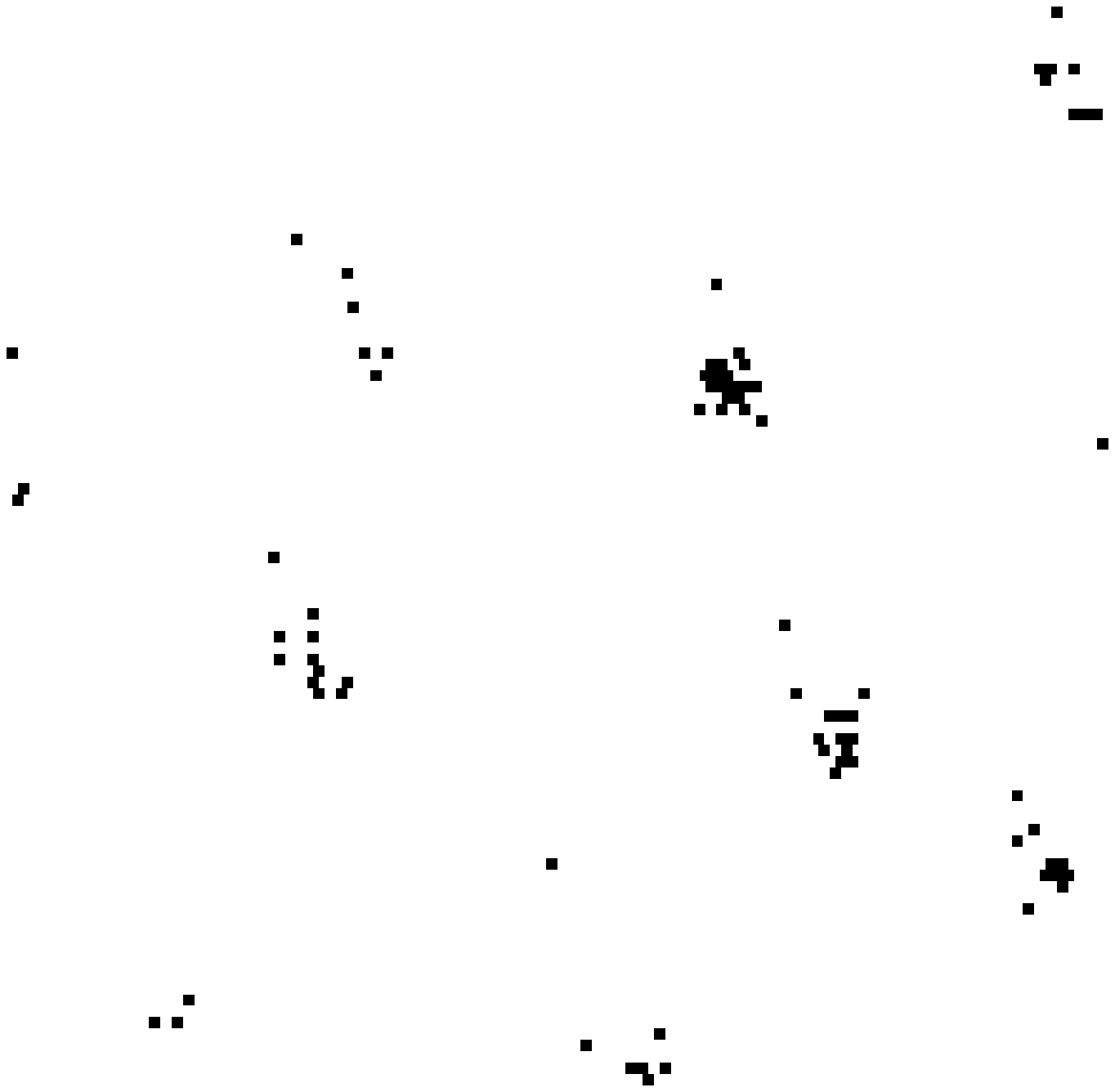,height=4.5cm,angle=0}} \hfill (b) \hfill
\fbox{\epsfig{figure=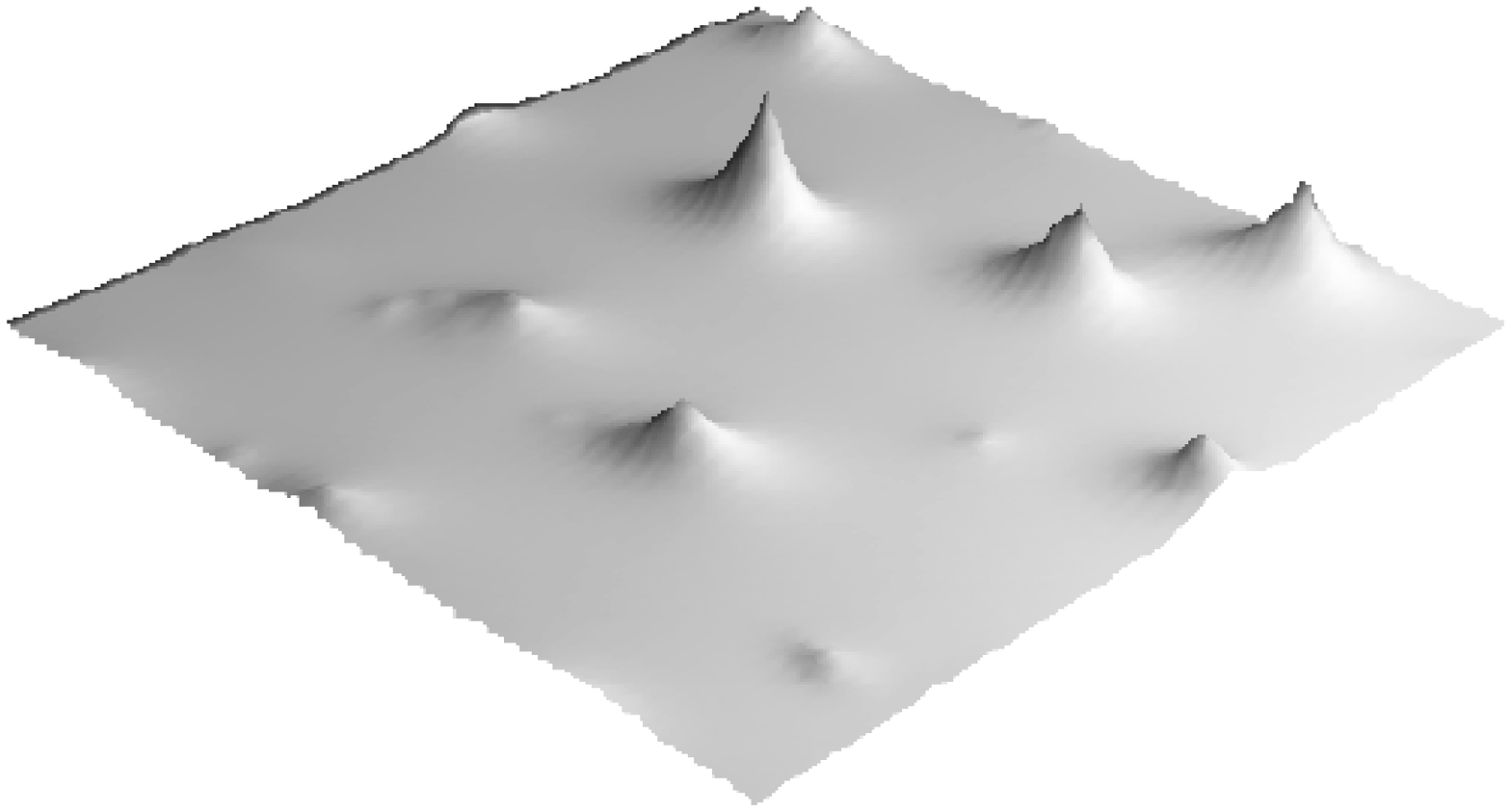,height=4.5cm,angle=0}}}\hfill
\vspace*{0.3cm}

\centerline{\hfill
\fbox{\epsfig{figure=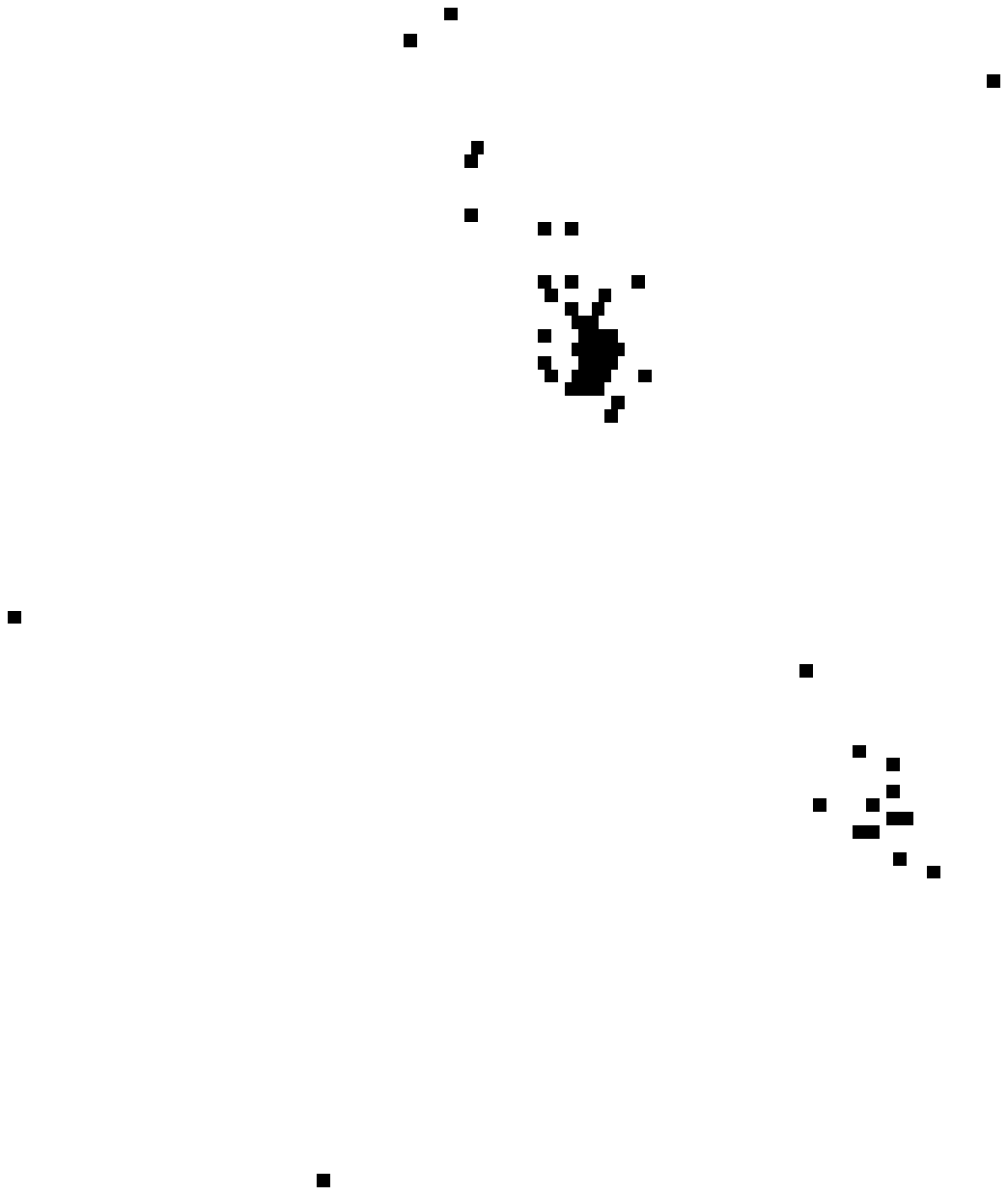,height=4.5cm,angle=0}} \hfill (c) \hfill
\fbox{\epsfig{figure=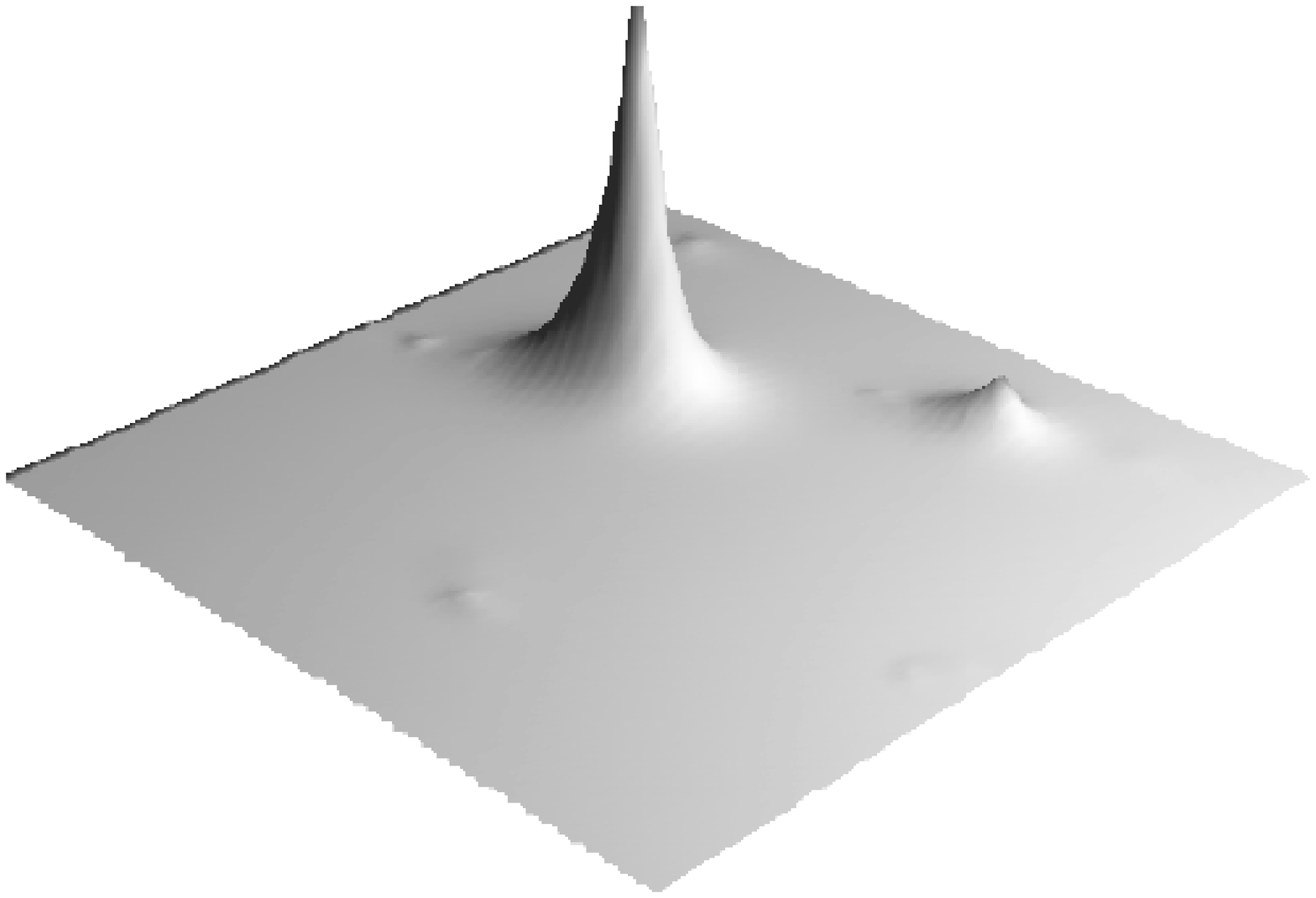,height=4.5cm,angle=0}}}\hfill
\vspace*{0.3cm}
\caption[]{
Snapshots of the positions of the active particles (left) and the
  distribution of the field (right) at different times: (a) $t=100$, (b) $t=5.000$, (c) $t=50.000$.
  (\textsc{Schweitzer} and \textsc{Schimansky-Geier} 1994)
 \label{compare}}
\end{figure}

\subsection{Formation of Trails}

A more complex dynamics of the particles can be obtained if instead of
the simple chemoattraction described above different chemical fields and
a more complex response of the particles to them are considered. So, let us
eventually assume that the active particles have another internal degree
of freedom $\theta_{i}$, in addition to their internal energy depot
$e_{i}$.  The individual parameter $\theta_{i}$ may be used to describe
different activities and responses to the field, i.e. the active
particles then become \emph{agents} with a more complex behavior
(\textsc{Schweitzer} 2002).

For example, the production rate of the field, $s$, may now depend on the
internal state $\theta_{i}\in \{-1,+1\}$, i.e. it becomes different for each
particle $i$ as follows:
\begin{equation}
  \begin{aligned}
s_i(\theta_i,t)=\frac{\theta_i}{2}\Big[&(1+\theta_i)\,s^{0}_{+1}\,
\exp\{-\beta_{+1}\,(t-t_{n+}^{i})\}\\  -\,&(1-\theta_i)\,s^{0}_{-1}\,
\exp\{-\beta_{-1}\,(t-t_{n-}^{i})\}\Big]
\label{prod}
\end{aligned}
\end{equation}
Eq. (\ref{prod}) means that the active particle, dependent on its
internal state $\theta_{i}$ may produce one of two different chemicals,
$\{+1\}$ or $\{-1\}$, with a rate that exponentially decreases in the
course of time.  Consequently, we now have two different chemical field
components that each are assumed to obey the following reaction equation
(diffusion is not considered here):
\begin{equation}
\frac{\partial h_{\theta}(\bbox{r},t)}{\partial t}=-k\,
h_{\theta}(\bbox{r},t) + \sum_{i=1}^{N}
s_i(\theta_i,t)\;\delta_{\theta;\theta_{i}}\;
\delta\Big(\bbox{r}-\bbox{r}_i(t)\Big) \;\;;\quad \theta \in \{-1,+1\}
\label{h-net-nd}
\end{equation}
The effect of the two field components on each active particle may be
described by an \emph{effective field}, that also depends on the internal
state $\theta_{i}$ of the agent, i.e. the gradient in \eqn{motion-h}
shall be replaced by the gradient of the effective field
(\textsc{Schweitzer} and \textsc{Tilch} 2002):
\begin{equation}
\frac{\partial h^{e}(\bbox{r},t)}{\partial \bbox{r}} =
\frac{\theta_i}{2}\,\left[\,
(1+\theta_i)\,\frac{\partial h_{-1}(\bbox{r},t)}{\partial \bbox{r}}\,-\,
(1-\theta_i)\,\frac{\partial h_{+1}(\bbox{r},t)}{\partial \bbox{r}}\,\right]
\label{dh-eff}
\end{equation}
The nonlinear feedback between the active particles and the chemical
field components can be summarized as follows: Particles with an internal
state $\theta_i=+1$ contribute to the field by producing component $+1$,
while they are affected by component $-1$, and particles with an internal
state $\theta_i=-1$ contribute to the field by producing component $-1$
and are affected by component $+1$.

Eventually, we assume that the particles can change their internal state
from $\theta_{i}=-1$ to $+1$ and vice versa, dependent on environmental
conditions or events. To be specific, we may consider that the active
particles are initially concentrated in a ``nest'' ($\theta_{i}=+1$) and
move out to search for ``food'', distributed in different spatial
locations. Once they found food, their initial state is changed to
$\theta_{i}=-1$, which means that the successful particles begin to
produce a different chemical (the ``success pheromone''). This gives a
new information to those particles that are not successful yet to find
the food sources, while the successful particle is guided back to the
nest by the already existing chemical field component $\{+1\}$.

As the result of this non-linear feedback between the active particles
communicating via two different chemical field components generated by
them, we can observe the formation of directed trails between a a nest
and different food sources (cf. Fig. 9)
\begin{figure}[htbp]
\centerline{\psfig{figure=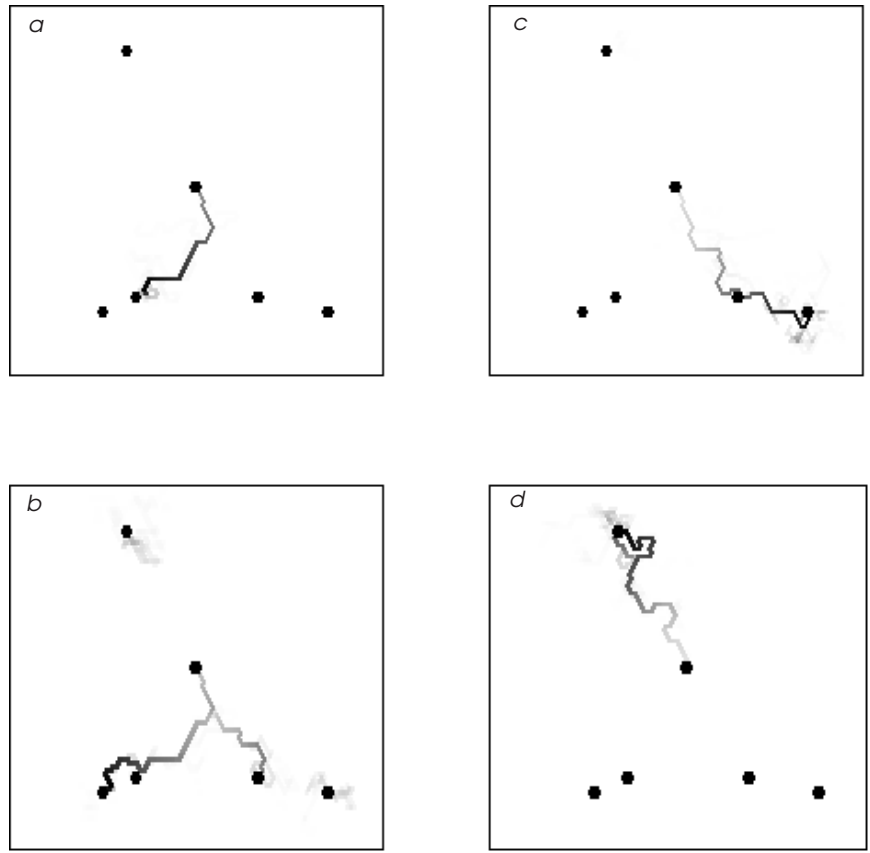,width=11.5cm}}
\caption[fig4]{
  Formation of trails from a nest (middle) to five randomly placed food
  clusters, which are assumed to be exhausted after a number of visits. The distribution of chemical component $\{-1\}$ (see text) is
  shown after (a) 2000, (b) 4000, (c) 8500, and (d) 15000 simulation time
  steps, respectively (\textsc{Schweitzer} \emph{et al.} 1997).
  \label{5food-sim}}
\end{figure}

We note that, with respect to biology, there are different parameters
which may influence trail following in addition to sensitivity, such as
trail fidelity, traffic density, detection distance, endurance of the
trail, navigation capabilities etc. (\textsc{Haefner} and \textsc{Crist}
1994, \textsc{Edelstein-Keshet} \emph{et al.}, 1995).  In contrast, our
model considers only minimal assumptions for the trail formation.  Here,
the formation of trail patterns is solely based on simple \emph{local}
chemical communication between the particles, with no additional
capabilities of orientation or navigation.  The spontaneous emergence of
a collective trail system by means of the active particles can be
described as a self-organizing process.  It turns out from the computer
simulations that, for different kinds of food sources, the model
generates a distinctive trail system to exploit the food sources, and it
performs a high flexibility in order to discover and to link new sources.

\section{Conclusions}

As the examples of the previous sections have shown, the approach of
active Brownian particles provides a suitable framework to consider both
the \emph{energetic conditions} for active motion and the
\emph{interactions} between the particles -- two ingredients essential
for active and coherent movement in biological systems.
The collective motion observed on the ``macroscopic'' level shows
interesting analogies to swarming phenomena found in flocks of bird,
schools of fish, but also in cells or insect societies.

With the established collective dynamics, we observe also the
\emph{emergence of new system properties} not readily predicted from the
basic equations. This process was, in the beginning of this paper,
described as \emph{self-organization}, i.e. ``the process by which
individual subunits achieve, through their cooperative interactions,
states characterized by new, emergent properties transcending the
properties of their constitutive parts.''  (\textsc{Biebricher} \emph{et
  al.} 1995). Whether or not these emergent properties occur, depends of
course, not only on the properties of the system elements and their
interactions, but -- as we have pointed out in Sect. 2 -- also on
suitable external conditions, such as global boundary conditions, the
in/outflux of resources (matter, energy, information).

For the prediction of the emergent properties from local interactions
fundamental limitations exist which are discussed, e.g., in chaos theory.
Moreover, stochastic fluctuations also give unlikely events a certain
chance to occur, which in turn affects the real history of the system.
This means, the properties of self-organizing systems cannot be
determined by a hierarchy of conditions, the system creates its
complexity in the course of evolution with respect to its global
constraints. Considering, that also the boundary conditions may evolve
and new degrees of freedom appear, co--evolutionary processes become
important, and the evolution may occur on a qualitatively new level.

Within our physical approach to these phenomena, we are basically
interested in the question \emph{which extensions} to a known (physical)
dynamics might bridge the gap towards a more complex (biological)
dynamics. Such a stepping stone strategy is quite promising, as various
applications for different biological problems have proven. Of course,
many details of real biological phenomena have necessarily to be dropped,
in order to focus on particular aspects.  Let us quote in this context
again from \textsc{Eigen's} foreword to the book of \textsc{Volkenstein}
(1994): ``The aim of theory is not to \emph{describe} reality in every
detail, but rather to \emph{understand} the principles that shape
reality.''

\section*{References}
\begin{description}

\item \textsc{Ben-Jacob, E., Schochet, O., Tenenbaum, A., Cohen, I.,
    Czir\'ok, A.,} and \textsc{Vicsek, T.}: Generic modelling of
  cooperative growth patterns in bacterial colonies.  \emph{Nature}
  \textbf{368}, 46--49 (1994)

\item \textsc{Biebricher, C.~K., Nicolis, G.,} and \textsc{Schuster, P.}:
  \emph{Self-Organization in the Physico-Chemical and Life Sciences},
  vol. 16546 of \emph{EU Report} (1995)

\item \textsc{Czirok, A., Ben-Jacob, E., Cohen, I.} and \textsc{Vicsek,
    T.}: Formation of complex bacterial colonies via self-generated
  vortices.  {\em Physical Review E\/} {\bf 54/2}, 1791--1801 (1996)

\item \textsc{Czirok, A., Barabasi, A.~L.} and \textsc{Vicsek, T.}:
  Collective motion of self-propelled particles: Kinetic phase transition
  in one dimension.  {\em Physical Review Letters\/} {\bf 82/1}, 209--212
  (1999)

\item \textsc{Deneubourg, J.~L., Gregoire, J.~C.,} and \textsc{Le~Fort,
    E.}: Kinetics of larval gregarious behavior in the bark beetle {\em
    Dendroctonus micans} (Coleoptera: Scolytidae).  \emph{J.  Insect
    Behavior} \textbf{3/2}, 169--182 (1990)

\item \textsc{Deutsch, A.}: Principles of morphogenetic motion: swarming
  and aggregation viewed as self-organization phenomena.  {\em J.
    Biosci.\/} {\bf 24/1}, 115--120 (1999)

\item \textsc{Ebeling, W.}, \textsc{Feistel, R.} and \textsc{Engel, A.:}
  Physik der Evolutionsprozesse, Berlin: Akademie-Verlag 1990

\item \textsc{Ebeling, W.} and \textsc{Feistel, R.}: Chaos und Kosmos -
  Prinzipien der Evolution, Heidelberg-Berlin-Oxford: Spektrum 1994

\item \textsc{Ebeling, W.}, \textsc{Freund, J.} and \textsc{Schweitzer,
    F.}: Entropie - Information - Komplexit\"at, Teubner:
  Stuttgart-Leipzig 1998

\item \textsc{Ebeling, W.}, \textsc{Schweitzer, F.} and \textsc{Tilch,
    B.}: Active Brownian motion with energy depots modelling animal
  mobility, \emph{BioSystems} \textbf{49}, 17-29 (1999)

\item \textsc{Ebeling, W.}, \textsc{Erdmann, U.}, \textsc{Dunkel, J.} and
  \textsc{Jenssen, M.}: Nonlinear dynamics and fluctuations of
  dissipative Toda chains, \emph{J. Stat. Phys.} \textbf{101}, 443-457
  (2000)

\item \textsc{Ebeling, W.} and \textsc{Schweitzer, F.}: Swarms of
  particle agents with harmonic interactions, \emph{Theory BioSci.}
  \textbf{120}, 1-18 (2001)

\item \textsc{Edelstein-Keshet, L., Watmough, J.,} and
  \textsc{Ermentrout, G.~B.}: Trail following in ants: individual
  properties determine population behaviour.  \emph{Behav. Ecol
    Sociobiol} \textbf{36}, 119--133 (1995)

\item \textsc{Erdmann, U.}, \textsc{Ebeling, W.}, \textsc{Schweitzer, F.}
  and \textsc{Schimansky-Geier, L.}: Brownian particles far from
  equilibrium, \emph{Europhys. J. B} \textbf{15}, 105-113 (2000)

\item \textsc{Haefner, J.~W.} and \textsc{Crist, T.~O.}: Spatial model of
  movement and foraging in harvester ants {\em (Pogonomyrmex)} (I): The
  Roles of Memory and Communication.  \emph{J. theor. Biol.}
  \textbf{166}, 299--313 (1994)

\item \textsc{H\"ofer, T.}: Chemotaxis and Aggregation in the Cellular
  Slime Mould.  In: M\"uller, S.C. Parisi, J. and Zimmermann, W. (Eds.),
  Transport and Structure. Their Competitive Roles in Biophysics
    and Chemistry, pp. 137--150, Berlin: Springer 1999

\item \textsc{Klimontovich, Yu. L.}: Statistical Physics of Open Systems,
  Dordrecht: Kluwer 1995

\item \textsc{Markl, H.}: Physik des Lebendigen, A.v.Humboldt-Magazin
  \textbf{65}, 13-24 (1995)

\item \textsc{Mikhailov, A.S.} and \textsc{Zanette, D.H.}: Noise-induced
  breakdown of coherent collective motion in swarms, Phys. Rev E
  \textbf{60}, 4571-4575 (1999)

\item \textsc{Okubo, M.} and \textsc{Levin, S.A.}: Diffusion and
  ecological problems: Modern perspectives, Berlin: Springer 2001

\item \textsc{Schienbein, M.} and \textsc{Gruler, H.}: Langevin equation,
  Fokker-Planck equation and cell migration, \emph{Bull. Math. Biol.}
  \textbf{55}, 585-608 (1993)

\item \textsc{Schweitzer, F.}: Brownian agents and active particles,
  Berlin: Springer 2002

\item \textsc{Schweitzer, F.}, \textsc{Ebeling, W.} and \textsc{Tilch,
    B.}: Complex motion of Brownian particles with energy depots,
  \emph{Phys. Rev. Lett.} \textbf{80}, 5044-5047 (1998)

\item \textsc{Schweitzer, F.}, \textsc{Ebeling, W.} and \textsc{Tilch,
    B.}: Statistical mechanics of canonical-dissipative systems and
  applications to swarm dynamics, \emph{Phys. Rev. E} \textbf{64},
  02110/1-12 (2001)

\item \textsc{Schweitzer, F., Lao, K.,} and \textsc{Family, F.}: Active
  random walkers simulate trunk trail formation by ants.
  \emph{BioSystems} \textbf{41}, 153--166 (1997)

\item \textsc{Schweitzer, F.} and \textsc{Schimansky-Geier, L.}:
  Clustering of active walkers in a two-component system.  \emph{Physica
    A} \textbf{206}, 359--379 (1994)
  
\item \textsc{Schweitzer, F.} and \textsc{Tilch, B.}: Self-assembling of
  network in an agent-based model. \emph{Physical Review E}
  \textbf{66}(2002) 026113 (1-9)

\item \textsc{Stevens, A.} and \textsc{Schweitzer, F.}: Aggregation
  Induced by Diffusing and Nondiffusing Media.  In: W.~Alt, A.~Deutsch,
  G.~Dunn (eds.), \emph{Dynamics of Cell and Tissue Motion}, pp.
  183--192, Basel: Birkh\"auser 1997

\item \textsc{Vicsek, T.}, \textsc{Czirok, A.}, \textsc{Ben-Jacob, E.},
  \textsc{Cohen, I.} and \textsc{Shochet, O.}: Novel type of phase
  transition in a system of self-driven particles, \emph{Phys. Rev.
    Lett.} \textbf{75}, 1226-1229 (1995)

\item \textsc{Vicsek, T.}: Fluctuations and scaling in biology, Oxford:
  University Press 2001

\item \textsc{Volkenstein, M.V.}: Physical approaches to biological
  evolution.  With a foreword by \textsc{M. Eigen}, Berlin: Springer 1994

\end{description}
\end{document}